\documentclass[prl,twocolumn,showpacs,preprintnumbers,aps,nofootinbib]{revtex4-2}

\usepackage{graphicx}
\usepackage{dcolumn}
\usepackage{bm}
\usepackage{amsmath,amssymb,amsfonts}
\usepackage{latexsym}

\usepackage{color}

\def\nn    {\nonumber}

\def\rbb{\rho_{bb}}
\def\rtt{\rho_{tt}}

\begin{document}


\title{\boldmath
Discovery Prospects for Electron and Neutron Electric Dipole Moments \\
 in the General Two Higgs Doublet Model}

\author{Wei-Shu Hou, Girish Kumar and Sven Teunissen}
\affiliation{Department of Physics, National Taiwan University, Taipei 10617, Taiwan}
%

 \date{\today}

\begin{abstract} 
Baryon asymmetry of the Universe offers one of the strongest hints for physics 
Beyond the Standard Model (BSM). Remarkably, in the {\it general} two 
Higgs Doublet Model ({\it g}2HDM) that possesses a second set of Yukawa 
matrices, one can have electroweak baryogenesis (EWBG) while the electron 
electric dipole moment (eEDM) is evaded by a {\it natural} flavor tuning that 
echoes SM. We show that eEDM may first emerge around $10^{-30}\,e$\,cm 
or so, followed by neutron EDM (nEDM) down to $10^{-27}\,e$\,cm. We 
illustrate a cancellation mechanism for nEDM itself, which in turn can be 
probed when a facility capable of pushing down to $10^{-28}\,e$\,cm 
becomes available.
\end{abstract}

\maketitle


\noindent{\it Introduction.---}
With no BSM physics emerging at the Large Hadron Collider (LHC), 
particle physics is in a state of exasperation. It is not clear whether one 
can address lofty issues such as the Baryon Asymmetry of the Universe, 
arguably one of the strongest hints for BSM physics that calls for the 
existence of {\it large} $CP$ violating (CPV) phase(s) beyond the 
Kobayashi-Maskawa phase~\cite{PDG} of SM.
The current frontier is the experimental race to measure electron EDM, 
where the bound held by the ACME experiment~\cite{ACME:2018yjb} has 
recently been surpassed at JILA~\cite{Roussy:2022cmp}, giving $d_e < 0.41 \times 
10^{-29}\,e$\,cm at 90\% C.L. This is several orders of magnitude stronger than the 
current nEDM bound of $d_n < 1.8 \times 10^{-26}\,e$\,cm by the nEDM experiment 
at PSI~\cite{Abel:2020pzs}. However, by using ultra cold neutrons (UCN), nEDM 
measurement is poised to improve by two orders of magnitude within two 
decades~\cite{Alarcon:2022ero}, with many experiments joining the fray.

In fact, the EDM experiments, much smaller than the behemoth LHC and its associated 
experiments, pose a {\it general} challenge: since BAU demands extremely large BSM 
CPV, can one survive the EDM bounds, especially eEDM? We explore this theme and 
promote the {\it general} two Higgs doublet model ({\it g}2HDM), where dropping 
the usual $Z_2$ symmetry one can have enough CPV for BAU, but the observed 
{\it flavor} (fermion mass and mixing) {\it hierarchies} --- a mystery in itself --- 
allows for an exquisite {\it natural flavor cancellation} mechanism to work for eEDM. 
We project that eEDM and nEDM could {well} emerge in the next decade 
or two, and extend the parameter range beyond previous considerations.

With {\it one} Higgs doublet observed, the two Higgs doublet 
model~\cite{Branco:2011iw} should be a no-brainer. A $Z_2$ symmetry is 
usually imposed to enforce the natural flavor conservation (NFC) condition 
posited by Glashow and Weinberg~\cite{Glashow:1976nt} to forbid extra 
Yukawa matrices of charged fermions. But as first illustrated by Cheng 
and Sher~\cite{Cheng:1987rs}, the flavor hierarchies may help alleviate 
Glashow's worries about flavor changing neutral couplings (FCNCs). It was 
pointed~\cite{Hou:1991un} out, even before the top discovery, that the 
process to watch, then, is $t \to ch$. The bound at the LHC, however, has 
reached the stringent ${\cal B}(t\,\to\,ch) < 0.00073$~\cite{CMS:2021hug}. 
But as stressed in 2013~\cite{Chen:2013qta} after the observation of $h(125)$, 
as the $\rho_{tc}$ coupling is associated more with the exotic $H$ and $A$ 
bosons, the $tch$ coupling should be $\rho_{tc}c_\gamma$, where 
$c_\gamma \equiv \cos\gamma$ is the $h$--$H$ mixing angle between the 
two $CP$-even scalars. 
{Who would have guessed that {\it Nature} would throw in, circa 2015, 
the {\it alignment} (small $c_\gamma$) phenomenon from the purely Higgs 
sector, to protect $t \to ch$ decay.}

Having introduced the $\rho_{tc}$ element of the up-type extra Yukawa matrix, it 
was subsequently shown~\cite{Fuyuto:2017ewj} that $\lambda_t\,{\rm Im}\,\rho_{tt}$ 
can robustly {drive EWBG~\cite{EWBG}}, with top Yukawa $\lambda_t \cong 1$ 
recently measured~\cite{PDG}, and with first order phase transition arising from 
$O(1)$~\cite{Kanemura:2004ch} Higgs quartic couplings, where there are a total of 7 
in absence of $Z_2$. It was further inferred {with emergent alignment that the exotic 
scalars are likely sub-TeV~\cite{Hou:2017hiw} in mass and populate 300--600~GeV}, 
openning up a search program at the 
LHC~\cite{Kohda:2017fkn, Ghosh:2019exx, Hou:2020chc, ATLAS:2023tlp}, 
where Ref.~\cite{ATLAS:2023tlp} is from ATLAS.

\begin{figure}[b!]
\center
\includegraphics[width=0.25 \textwidth]{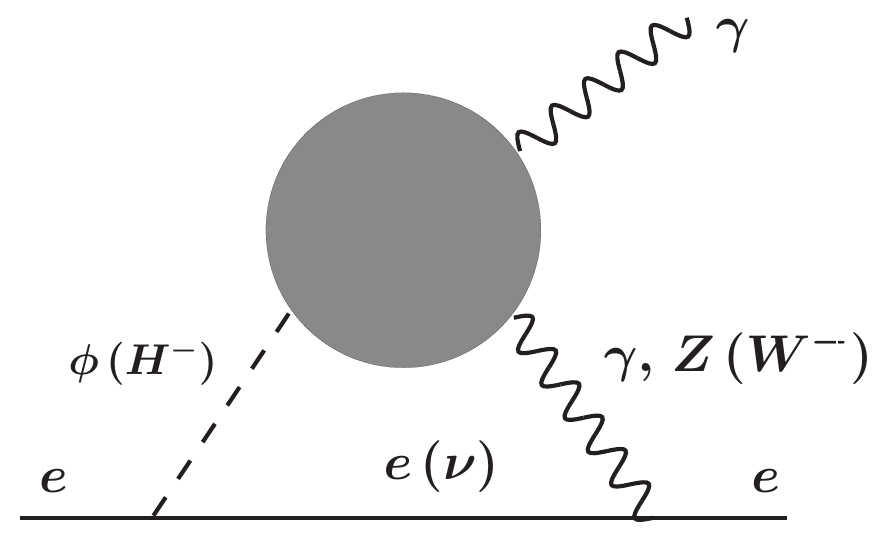}
\vskip-0.1cm
\caption{
 A two-loop Barr-Zee diagram for electron EDM with extra Yukawa coupling
 $\rho_{ee}$ on electron line, and top (hence $\rho_{tt}$) and $W$ run in the
 gray blob for neutral scalar $\phi = h, H, A$. Neutron EDM has many more
 contributions, including $u$- and $d$-quark chromo-moments and
 the Weinberg operator.
}
\label{fig:Barr-Zee}
\end{figure}

\begin{figure*}[t!]
\center
\includegraphics[width=6.5cm,height=4.5cm]{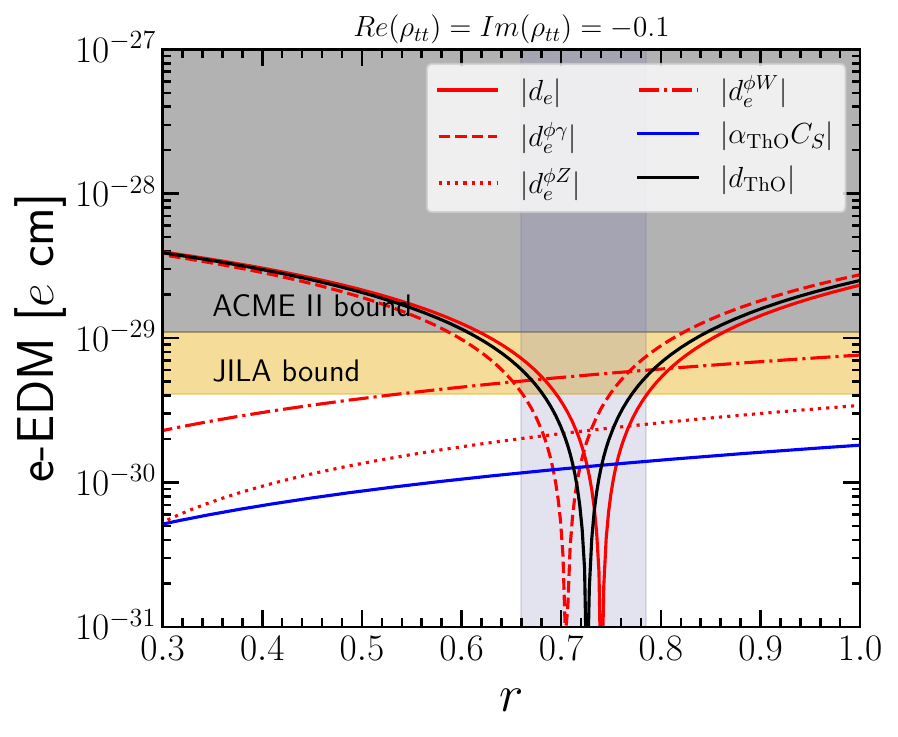}\includegraphics[width=5.6cm,height=4.5cm]{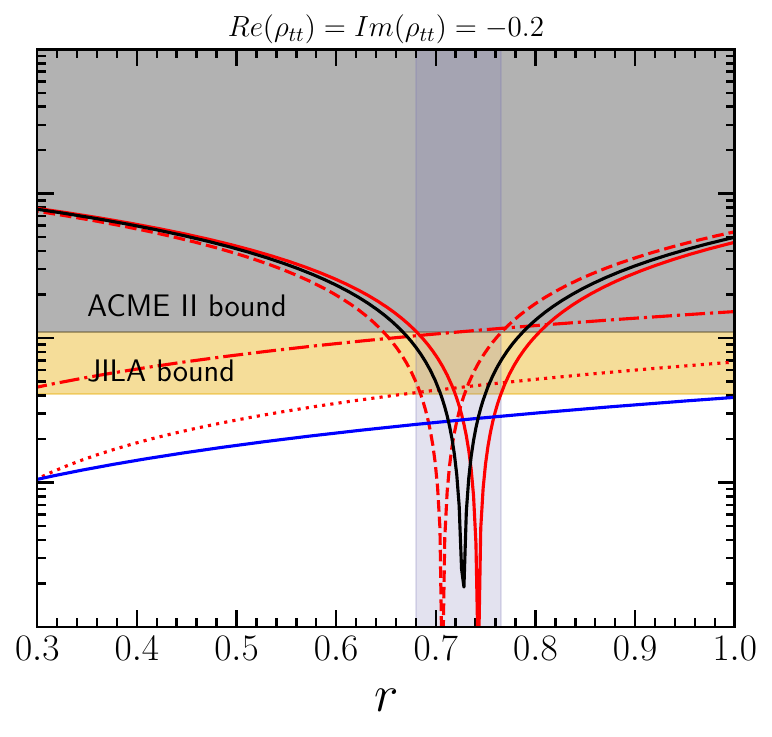}\includegraphics[width=5.6cm,height=4.5cm]{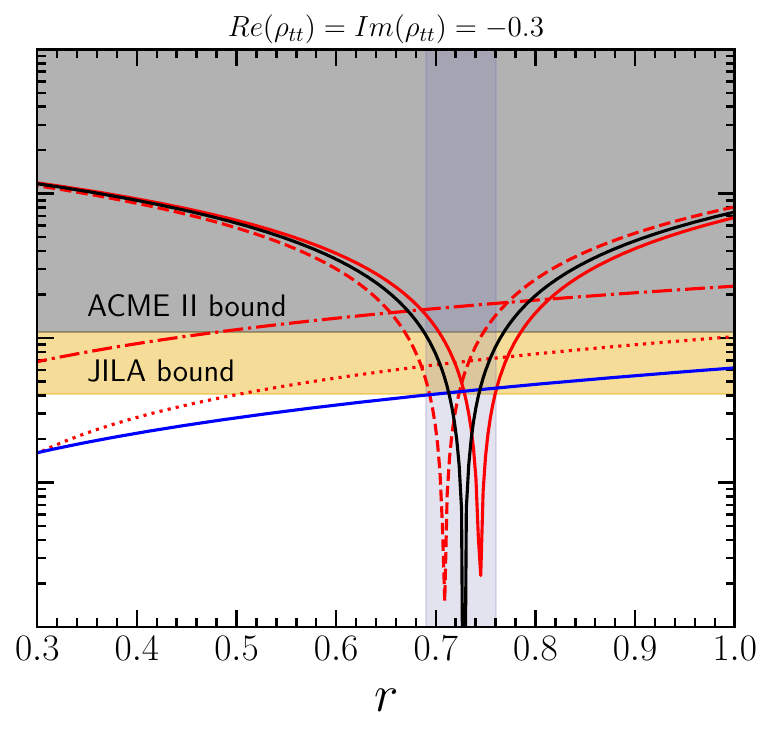}
\vskip-0.1cm
\caption{
 Cancellation mechanism~\cite{Fuyuto:2019svr} for eEDM for $\rho_{tt}$ values
 in Eq.~(4), $c_\gamma = 0.1$, and $m_H, m_A, m_{H^+} = 500$ GeV.
}
\label{fig:eEDM}
\end{figure*}

The large ${\rm Im}\,\rho_{tt}$ at $O(\lambda_t) \sim 1$ that drives EWBG 
brings up our theme of how to survive eEDM. A typical two-loop Barr-Zee
 diagram~\cite{Barr:1990vd} for eEDM is given in Fig.~\ref{fig:Barr-Zee}. 
To cancel the leading effect due to $\rho_{tt}$ {\it and} $\rho_{ee}$, specifically 
the $\phi\gamma\gamma^*$ insertion, one finds~\cite{Fuyuto:2019svr} 
\begin{align}
   |\rho_{ee}/\rho_{tt}| = r|\lambda_e/\lambda_t|,
   \ \ \ \ \arg(\rho_{ee}\rho_{tt}) = 0,
\label{cancel}
\end{align}
with $r \simeq 0.7$, where the first relation follows from a phase-lock between 
$\rho_{ee}$ and $\rho_{tt}$ for $\phi = A$. Eq.~(1) is remarkable in that the 
$\rho$ matrices seem to ``know'' the quark mass and mixing hierarchies in SM.

The purpose of this Letter is to show that the combined eEDM and nEDM effort provides 
the cutting edge probe of $\rho_{tt}$-driven EWBG in {\it g}2HDM: as the experimental 
competition heats up, we may first observe eEDM in the $10^{-30}$--$10^{-31}\,e$\,cm range, 
followed by confirmation at n2EDM at PSI for $d_n \sim 10^{-26}$--$10^{-27}\,e$\,cm in 
about a decade. But as we will illustrate a general cancellation mechanism for nEDM itself, 
a more advanced nEDM experiment may confirm down to $10^{-28}\,e$\,cm in two decades.
To unravel the underlying dynamics, the ``decadal mission''\;\cite{Hou:2021wjj} 
with direct exotic scalar search at the LHC, flavor physics explorations with LHCb and 
Belle II, plus $\mu$ and $\tau$ studies, would be needed.

\vskip0.2cm
\noindent{\it g2HDM and EDMs.---}
For simplicity, we assume $CP$-conserving~\cite{Hou:2017hiw, Davidson:2005cw} 
Higgs potential of g2HDM, removing it as a CPV source without discussing it any
further here, so CPV is relegated to extra Yukawa couplings.
As already stated, $O(1)$ Higgs quartics supply~\cite{Fuyuto:2017ewj,
 Kanemura:2004ch} the prerequisite first order EW phase transition for BAU, 
which is a bonus in {\it g}2HDM.

To clarify the flavor and EWBG discussion in the Introduction, without any 
$Z_2$ symmetry, there are extra Yukawa matrices $\rho^f$ for charged fermions 
$f = u,\,d,\,\ell$~\cite{Davidson:2005cw, Hou:2020chc}, which are complex and 
nondiagonal,
\begin{align}
\mathcal{L} = 
 - & \frac{1}{\sqrt{2}} \sum_{f = u, d, \ell}
 \bar f_{i} \Big[\big(-\lambda^f_i \delta_{ij} s_\gamma + \rho^f_{ij} c_\gamma\big) h \nn\\
  & + \big(\lambda^f_i \delta_{ij} c_\gamma + \rho^f_{ij} s_\gamma\big)H
    - i\,{\rm sgn}(Q_f) \rho^f_{ij} A\Big]  R\, f_{j} \nn\\
 - & \bar{u}_i\left[(V\rho^d)_{ij} R-(\rho^{u\dagger}V)_{ij} L\right]d_j H^+ \nn\\
 - & \bar{\nu}_i\rho^L_{ij} R \, \ell_j H^+ + {\rm h.c.},
\label{eff}
\end{align}
with generation indices $i$, $j$ summed over, $L, R = 1\mp \gamma_5$, and $s_\gamma 
\equiv \sin\gamma$. The $A$, $H^+$ couplings are $c_\gamma$-independent, while in 
the alignment limit ($c_\gamma \to 0$, $s_\gamma \to -1$), $h$ couples diagonally and 
$H$ couples via extra Yukawa couplings $-\rho_{ij}^f$, which can drive BAU. Thus, 
besides mass-mixing hierarchy protection~\cite{Hou:1991un} of FCNCs, alignment 
provides~\cite{Hou:2017hiw} further safeguard, such as for $t\to ch$, without the need 
of NFC. Furthermore, the $\mu^2_{12}\Phi^\dag\Phi'$ term in the Higgs potential is 
eliminated after symmetry breaking by minimization, leaving a unique $h$-$H$ mixing 
parameter, $\eta_6$, which can be $O(1)$~\cite{Hou:2017hiw} for small $c_\gamma$, 
with $H$, $A$, $H^+$ likely in the 300--600 GeV mass range.

Considering how  effective {\it g}2HDM {\it evades} stringent flavor constraints, 
and to address the question ``What makes {\it g}2HDM so well hidden so far?'',
we guessed a ``rule of thumb''\,\cite{Hou:2020itz} for flavor control:
\begin{align}
   \rho_{ii} \lesssim {\cal O}(\lambda_i),\;\;
   \rho_{1i} \lesssim {\cal O}(\lambda_1),\;\;
   \rho_{3j} \lesssim {\cal O}(\lambda_3),
\label{rho_ij}
\end{align}
{with} $j \neq 1$. This allows $\rtt = {\cal O}(1)$ but $\rbb \simeq 0.02$. However, 
$\rho^d_{ij}$ seems to be an order of magnitude weaker by flavor constraints. 

{With complications of transport equations for EWBG\,\cite{Fuyuto:2017ewj}, 
the simplified case with $H,\,A,\,H^+$ degenerate at 500~GeV was studied.} 
The ACME experiment~\cite{ACME:2018yjb} taught us the lesson to keep 
the weakest $\rho_{ee}$ coupling in the Barr-Zee diagrams of Fig.~1, where the 
exquisite cancellation mechanism of Eq.~(1) was uncovered~\cite{Fuyuto:2019svr}. 
The prowess of ACME, however, led one to illustrate with the 
timid $|\rho_{tt}| \simeq 0.1$, which we seek to extend here.

\vskip0.15cm
\noindent{\it Results: Interplay of eEDM and nEDM.---}
In our numerical illustration, we shall keep the degeneracy at 500~GeV, 
but explore a broader range of 
\begin{align}
 {\rm Re}\,\rho_{tt} =  {\rm Im}\,\rho_{tt}  = -0.1, -0.2, -0.3,
\label{rho_tt-range}
\end{align}
and follow the numeric ansatz~\cite{Fuyuto:2019svr} for $f = u, c; d, s, b$,
\begin{align}
{\rm Re}\,\rho_{ff} = -r\frac{\lambda_{f}}{\lambda_t}{\rm Re}\,\rho_{tt},\quad
 {\rm Im}\,\rho_{ff} = +r\frac{\lambda_{f}}{\lambda_t}{\rm Im}\,\rho_{tt},
\label{ansatz}
\end{align}
where $r \simeq 0.71$ is a combination of loop functions that is 
insensitive~\cite{Fuyuto:2019svr} to exotic Higgs spectrum.

\begin{figure}[t!]
\center
\includegraphics[width=0.48 \textwidth]{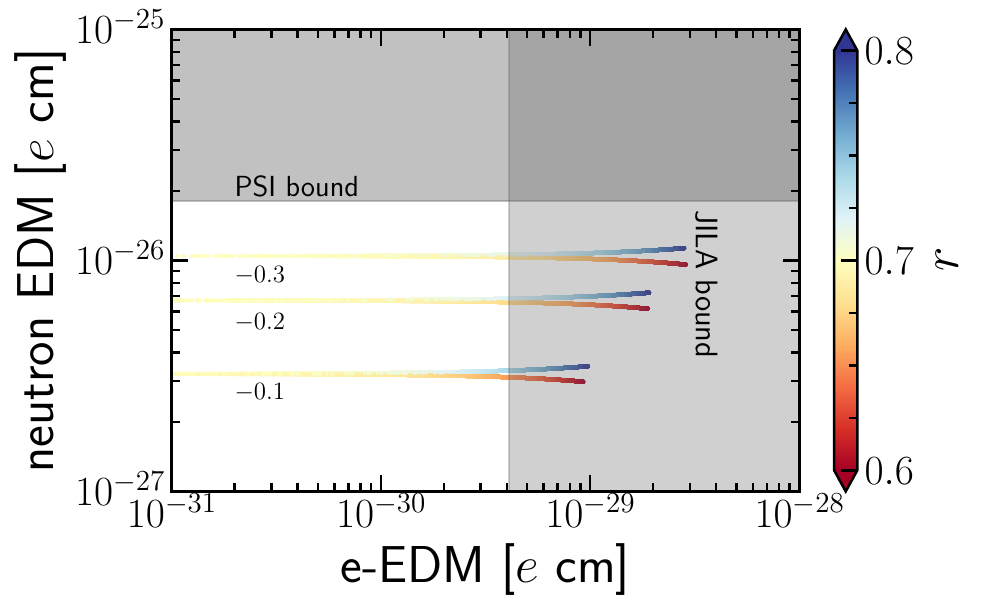}
\vskip-0.1cm
\caption{
 Combined scan result for $r \in [0.6, 0.8]$ for electron and
 neutron EDM for same range of $\rho_{tt}$ and exotic Higgs
 masses as in Fig.~2, with $\rho_{ff}$ fixed according to Eq. (5).
}
\label{fig:eEDM-nEDM}
\end{figure}

In Fig.~\ref{fig:eEDM} we illustrate the {\it natural} ``flavor tuning''~\cite{Fuyuto:2019svr} 
of Eq.~(1), for $\rho_{tt}$ values in Eq.~(4) and numeric ansatz of Eq.~(5), where both 
bounds of ACME~\cite{ACME:2018yjb} and JILA~\cite{Roussy:2022cmp} are shown. We 
{take} some liberty in the visual effect of the light purple band, with left side taken from the 
red-dashed $d_e^{\phi\gamma}$ curve~\cite{Fuyuto:2019svr}, and right side from the 
red-solid $d_e$ curve. This is in part because, though the cancellation point (black-solid 
curve sitting in the middle, with final shift from $C_S$ effect~\cite{Fuyuto:2019svr}) is 
insensitive to the spectrum~\cite{Fuyuto:2019svr}, there should be some spread in exotic 
scalar masses, which we refrain from exploring.

From left to right in Fig.~2, as $\rho_{tt}$ strength rises, the ``funnel'' is raised, but 
at $10^{-30}\,e$\,cm, the openning of the funnel is still decent, suggesting a still 
robust discovery likelihood, although by $10^{-31}\,e$\,cm, it approaches a pinpoint 
and may no longer seem plausible. In any case, these plots are for numeric illustration.

Turning to nEDM, besides effects of $\rho_{uu}$ and $\rho_{dd}$ through
Barr-Zee type diagrams, there are also chromo-moments and the Weinberg
operator, with progressively larger theory uncertainties. While the classic review 
of Pospelov and Ritz~\cite{Pospelov:2005pr} continue to be widely cited, it is a 
bit dated. We use the more recent formula~\cite{Hisano:2015rna},
\begin{align}
 d_n = - 0.20\,d_u + 0.78\,d_d & + e\,(0.29\,\tilde d_u + 0.59 \tilde d_d) \nn \\
                         & + e\,23\;{\rm MeV}\,C_W,
\label{d_n}
\end{align}
{where we evaluate chromo-moments $\tilde d_{u, d}$ and the Weinberg 
operator $C_W$ term~\cite{Kaneta:2023wrl} by following Refs.~\cite{Abe:2013qla} 
and~\cite{Jung:2013hka}, respectively. 
A recent discussion on uncertainties can be found in Ref.~\cite{Kaneta:2023wrl}.}

We give in Fig.~\ref{fig:eEDM-nEDM} the scan plot for $r \in [0.6, 0.8]$ for same 
range of $\rho_{tt}$ and exotic Higgs masses as in Fig.~2, showing both the JILA 
bound~\cite{Roussy:2022cmp} on eEDM, and PSI bound~\cite{Abel:2020pzs} on 
nEDM. One survives the PSI bound even for $|\rho_{tt}| \simeq 0.3\sqrt2$, while 
$r \simeq 0.7$ nicely illustrates the {\it natural} flavor cancellation of eEDM.
The follow-up experiment to nEDM at PSI, i.e. n2EDM~\cite{n2EDM:2021yah}, 
plans to reach down to $10^{-27}\,e$\,cm sensitivity within a decade, and should 
be able to cover the range illustrated in Fig.~3.

\begin{figure*}[t!]
\center
\includegraphics[width=5.9cm,height=4.5cm]{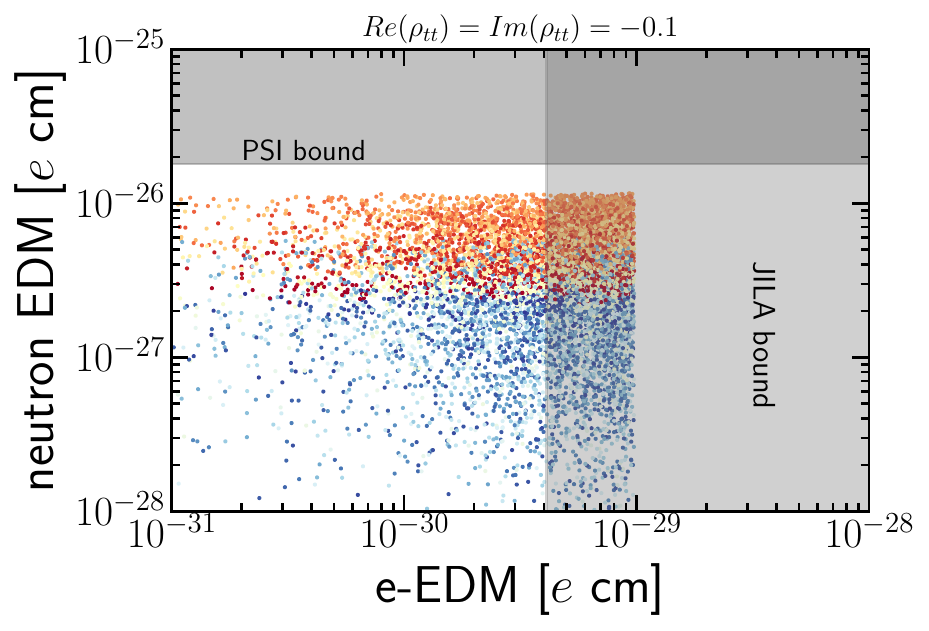}
\includegraphics[width=5.192cm,height=4.5cm]{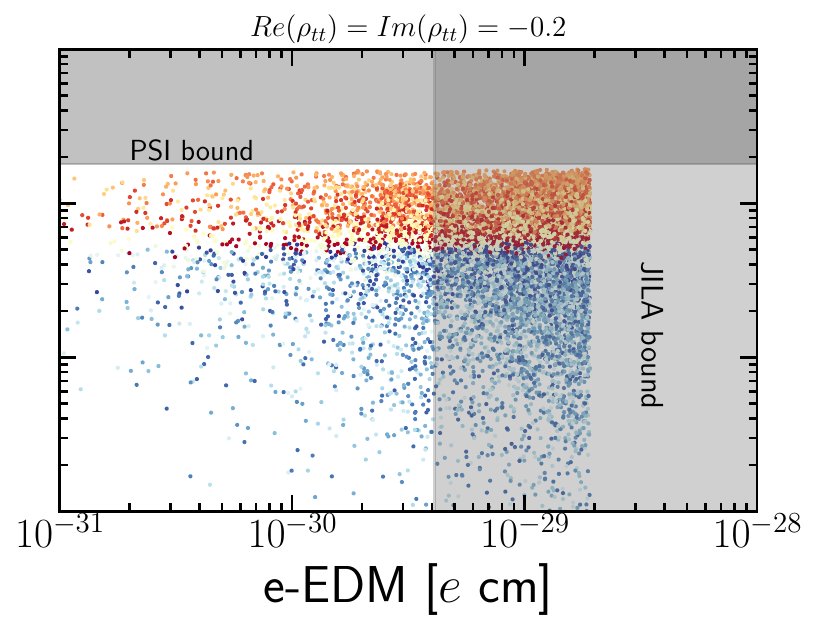}
\includegraphics[width=6.352cm,height=4.5cm]{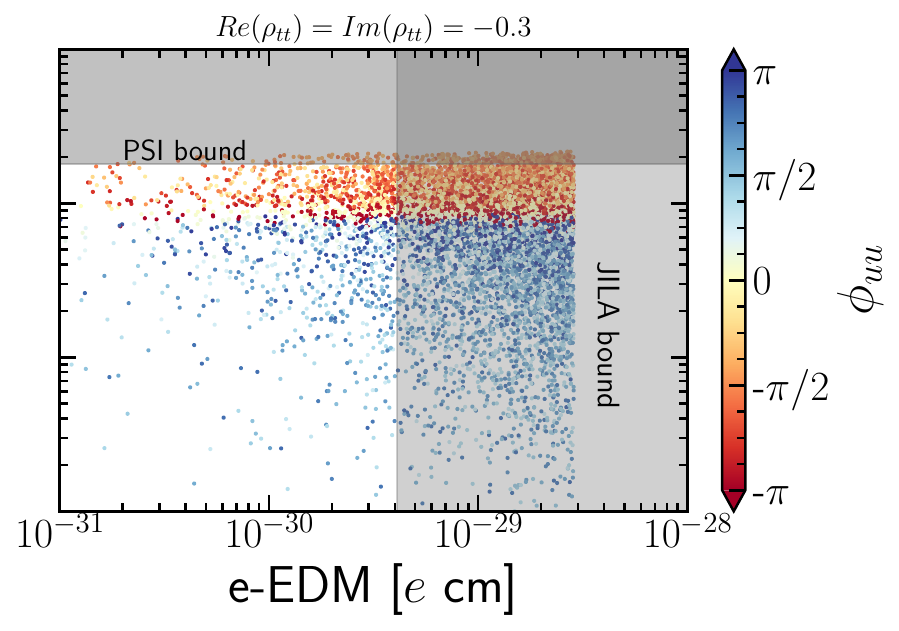}
\vskip-0.1cm
\caption{
 Scan for eEDM and nEDM using Eq. (1) as in Fig. 3,\,except varying
 $|\rho_{uu}|\,\in\,[0.3\lambda_u,\,3\lambda_u]$, $\phi_{uu}\,\in\,[-\pi,\,\pi]$.
}
\label{fig:eEDM-nEDM-var-u}
\end{figure*}

But we should admit that Eq.~(5) is nothing but an ansatz~\cite{Fuyuto:2019svr} 
for sake of numeric illustration. The fact is, we have little knowledge of the actual 
strength of extra Yukawa couplings such as $\rho_{uu}$. Our ``rule of thumb'' of 
Eq.~(3) is our guess of the ``flavor protection'' in {\it g}2HDM, which echos the 
remarkable cancellation mechanism of Eq.~(1) for eEDM. Taking Eq.~(3) literally, 
it states that {$|\rho_{uu}| = O(\lambda_u)$, with phase unknown.} 
Thus, {taking the usual sense of ``an order of magnitude''}, we vary
\begin{align}
 |\rho_{uu}| \in [0.3\lambda_u, 3\lambda_u], \quad \arg\rho_{uu} \in [-\pi, \pi],
\label{var-u}
\end{align}
while keeping other $\rho_{ff}$s according to Eq.~(5). This explores the 
impact of $\rho_{uu}$ strength and phase on nEDM.
Since $\rho_{tt}$ is in the 3rd quadrant in Eq.~(4), in the convention of Eq.~(7), 
$\arg\rho_{tt} = -3\pi/4$.

A scan plot of the variation of Eq.~(7) is given in Fig.\;\ref{fig:eEDM-nEDM-var-u} 
for illustration. For negative $\arg\rho_{uu}$, 
nEDM is closer to the PSI bound (red and yellow scan points), and for the largest 
$|\rho_{tt}| = 0.3\sqrt2$ (right plot), the bound cuts {a little bit} into the scan space. 
But interestingly, for positive $\arg\rho_{uu}$, i.e. {\it opposite} the sign of $\arg\rho_{tt}$, 
the blue scan points extend below $10^{-27}\,e$\,cm, which can evade n2EDM of PSI. 
Therefore, the scan in Fig.~4 illustrates a {\it general} cancellation mechanism that 
may well be operative in {\it Nature} for neutron EDM. It can be probed, however, at 
more advanced nEDM facilities, such as the nEDM experiment under construction at 
the Spallation Neutron Source~\cite{nEDM@SNS-ORNL} at Oak Ridge National Lab 
(ORNL), which utilizes UCN and can probe down to $10^{-28}\,e$\,cm. Although 
this may go beyond the next decade, the possibility appears to be covered fully, as the 
blue scan points tend to run out by $10^{-28}\,e$\,cm.

Thus, if {\it g}2HDM is the source of EWBG, the combined effort of eEDM and 
nEDM experiments seem poised for major discoveries in the coming decade or two.

\vskip0.2cm
\noindent{\it Discussion and Summary.---}
This work was actually stimulated by the ability {at the LHC} to probe 
top CPV, i.e. top chromo-moments~\cite{CMS-tCPV}. As this is a new 
beginning, {top chromo-moment bounds are still} rather weak. We 
realized instead that prospects for electron and neutron EDMs are 
rather good in {\it g}2HDM.

We have kept $H$, $A$ and $H^+$ degenerate at 500~GeV and have not
revisited EWBG, but we have checked that features at 300~GeV are quite 
similar, where baryogenesis should be more efficient. The actual parameter 
space should therefore be considerably larger. For example, breaking the 
degeneracy, one would need to face precision electroweak constraints~\cite{PDG}, 
where either one keeps $m_A = m_{H^+}$ (custodial symmetry), or take 
the twisted-custodial~\cite{Gerard:2007kn} case of $m_H = m_{H^+}$.

We have emphasized {as our theme} that it is nontrivial that {\it g}2HDM 
can provide electroweak baryogenesis while surviving the eEDM constraint, 
a remarkable feat rooted in the {\it flavor} structure as revealed by the 
SM sector. With exotic $H$, $A$ and $H^+$ bosons sub-TeV in mass, 
search programs at the LHC~\cite{ATLAS:2023tlp} have started, while 
there are also some good flavor probes~\cite{Hou:2020itz}.
Any BSM theory {of EWBG} would need to face the litmus test of 
surviving the eEDM bound~\cite{Roussy:2022cmp}.

We may sound optimistic in the discovery prospect for eEDM at $10^{-30}\,e$\,cm. 
Note that both the JILA and ACME bounds are still consistent {even with} 
$O(10^{-29})\,e$\,cm. Considering possible fluctuations in data, discovery not far 
below the existing bound is quite plausible, especially if {\it Nature} has already 
marked {\it g}2HDM up for baryogenesis. {A known example} is the ARGUS 
discovery~\cite{ARGUS:1987xtv} of $B^0$--$\bar B^0$ mixing, which 
practically sits right on top the previous CLEO~\cite{CLEO:1986yet} bound.

In summary, {\it g}2HDM without $Z_2$ symmetry achieves baryogenesis 
but can evade the eEDM bound by {\it natural} flavor tuning. 
Electron EDM may harbinger  a new era, echoed not long after by neutron 
EDM; while this does not prove {\it g}2HDM is behind EWBG, it would
likely become a frontrunner.
With exotic Higgs search at the LHC, ongoing efforts at Belle~II and other 
flavor fronts, and with excellent prospects for electron and neutron EDM 
measurements, the future looks bright for unveiling what may actually lie 
behind baryogenesis.

\vskip0.1cm
\noindent{\bf Acknowledgments} \
We thank the support of grants NSTC 112-2639-M-002-006-ASP, 
and NTU 112L104019 and 112L893601.


\end{document}